# A program for SAXS data processing and analysis

Li Zhi-hong(李志宏)[1)]

Beijing Synchrotron Radiation Facility, Institute of High Energy Physics, Chinese Academy of Sciences, Beijing 100049, China

Author email: box@china.com.cn.

**Abstract** A computer program for small angle X-ray scattering (SAXS) data processing and analysis named S.exe written in Intel Visual Fortran has been developed. This paper briefly introduces its main theory and function.

## I. Introduction

Small angle X-ray scattering (SAXS) is a powerful analytical method for determining the nanoscale structure of matter [1] based on the detection of X-rays scattered by the sample at very low angles using dedicated instruments. SAXS is applicable to any system exhibiting fluctuations in electron density, such as biopolymers in aqueous dispersions, nanoparticles in a solid or liquid matrix or nanopores in a solid matrix. Computer programs play a key role in SAXS data analysis [2]. We have developed a program named S.exe written in Intel Visual Fortran for SAXS data processing and analysis. This short paper briefly introduces the main theory and function of the program.

## II. Main theory

The discussion below is based on point-collimation (pinhole) scattering data, though the program can also be used for slit-smeared data. For slit smeared data, one can perform desmearing [1] or use the data directly with special formulae [3].

## 1 Primary data processing

The primary data processing refers mostly to scattering angle (or scattering vector) calibration, normalization to the incident beam intensity and background subtraction. The scattering angle or scattering vector calibration is performed by using the knowledge of the incident X-ray wavelength and the sample to detector distance, or by using the diffraction data of a standard sample. It should be noted that before the calibration, the raw scattering profile recorded directly by a 1D detector or derived from an image of a 2D detector must be preprocessed to check and correct the position of the centre of the direct beam, determine the radial distance on the scattering profile and choose the useful data. Normalization and background subtraction are done with:

$$I(q) = I_s(q) - \frac{K_s}{K_b} I_b(q) \qquad (1)$$

where $I(q)$ is the background subtracted scattering intensity of the sample; $I_s(q)$ and $I_b(q)$ are the raw scattering intensity of sample and background, respectively; $K_s$ and $K_b$ are the intensity of the beam transmitted through the sample and the background respectively [4]. Besides, the desmearing of the slit-collimation data is also optional [1,5].

## 2 Data analysis

### 1) Porod analysis

Porod's law is one of the basic formulae in SAXS [6,7], which mainly describes the asymptotic behavior of the intensity as a function of scattering vector. The three situations of scattering characteristics for different sample structures are illustrated in the Porod plot in Figure 1.

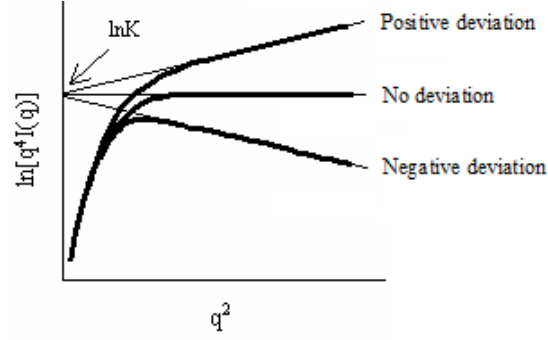

Figure 1  Porod plot with point-collimation in SAXS

**(1) No deviation**

It is well known that only the scattering by an ideal two-phase system with sharp boundary and with constant but different electron density in each phase obeys the Porod's law [6,7]:

$$\lim q^4 I(q) = K \quad \text{or} \quad \lim \ln[q^4 I(q)] = \ln K \tag{2}$$

where q is the scattering vector, $q=4\pi\sin\theta/\lambda$, $2\theta$ is the scattering angle, $\lambda$ is the incident X-ray wavelength, I(q) is the point-collimation (pinhole) scattering intensity, K is the Porod constant. Equation (1) means that the plot of $\ln[q^4 I(q)]$ vs $q^2$ at high q-values tends to a constant as the middle curve in Figure 1 illustrates.

The specific surface (total surface per unit of volume) of the sample can be calculated by the Porod method [8]:

$$S_V = \pi P(1-P)\frac{K}{Q} \tag{3}$$

where K is the Porod constant; P is the volume fraction of the dispersed phase, for porous systems, P is called porosity, and it is usually derived from the densities of the sample or the absolute intensity of the primary beam; Q is the invariant, which is given by [8]:

$$Q = \int_0^\infty q^2 I(q) dq \tag{4}$$

Q should be calculated by extrapolating scattering intensity towards zero and infinitely large scattering angle, i.e., into regions where no experimental data are available. The first extrapolation can be done by applying the Guinier approximation and the second extrapolation can be done by applying Porod's $q^{-4}$ dependence.

**(2) Negative deviation**

For a quasi two-phase system with a diffuse phase boundary or a transition zone (i.e. an interface layer) with average thickness E between two phases, there is a reduction of scattering especially at high angles and this results in a negative deviation from the Porod's law (see Figure 1). A factor as $\exp(-\sigma^2 q^2)$, which is the Fourier transform of the autocorrelation of the Gaussian smoothing function for a sigmoidal–gradient model, is used to represent the negative deviation due to the diffuse interface. $\sigma$ is the standard deviation of the Gaussian smoothing function which is a parameter related to the thickness of the interfacial layer. Then, for the negative deviation from the Porod's law, one obtains [7,9]:

$$I(q) = \frac{K}{q^4}\exp(-\sigma^2 q^2) \quad \text{or} \quad \ln[q^4 I(q)] = \ln K - \sigma^2 q^2 \qquad (5)$$

Using this equation to fit the negative deviation from the Porod's law, one can derive K and $\sigma$. The average thickness $E$ of the interface layer is then given by [7,9]:

$$E = (2\pi)^{1/2}\sigma \qquad (6)$$

**(3) Positive deviation**

For a quasi two-phase system having sharply defined phase boundaries but with micro-fluctuations of electron density within any phase of a two-phase system, there is some additional scattering and this results in a positive deviation from the Porod's law (see Figure 1) [6,10,11,12]. A factor like $\exp(bq^2)$ might be used to represent the positive deviation, where b is a constant related to the size of the regions with micro-fluctuations of electron density. The actual relationship remains to be further studied. Then, for the positive deviation from the Porod's law, one obtains [12]:

$$I(q) = \frac{K}{q^4}\exp(bq^2) \quad \text{or} \quad \ln[q^4 I(q)] = \ln K + bq^2 \tag{7}$$

**(4) Correction of deviation**

Once the value of σ is derived from the negative deviation analysis, the correction of negative deviation from the Porod's law can be performed [9]:

$$I'(q) = \exp(\sigma^2 q^2) I(q) \tag{8}$$

Similarly, once the value of b is obtained from the positive deviation analysis, the correction of positive deviation from the Porod's law can be performed [12]:

$$I'(q) = \exp(-bq^2) I(q) \tag{9}$$

**2) Debye analysis**

The correlation function γ(r) is often used for statistical analysis in SAXS [13,14]. It is the Fourier transform of the scattering intensity I(q) and may be analyzed in terms of an ideal lamellar morphology [1] to obtain structural parameters describing the sample.

$$\gamma(r) = \frac{1}{2\pi \overline{(\Delta\rho)^2} V} \int_0^\infty q^2 I(q) \frac{\sin qr}{qr} dq \tag{10}$$

where $\overline{(\Delta\rho)^2}$ is the mean square value of electron density difference; V is the X-ray irradiated sample volume. The function γ(r) describes the correlation of electron density function between any two points with distance r within a system, and determines the distribution of the scattered intensities. The specific surface $S_V$ of any two phases system can also be derived from γ(r) [1,8]:

$$S_V = -4P(1-P)\left[\frac{d\gamma(r)}{dr}\right]_{r=0} \tag{11}$$

**(1) No deviation**

For an ideal two-phase system with sharp boundary, γ(r) has exponential form

[14,15]:

$$\gamma(r) = \exp(-r/A_c) \tag{12}$$

where $A_c$ is called the correlation distance (long range order, distances between similar structures) which is a measure of the phase size. According to the Debye scattering theory [14-17]:

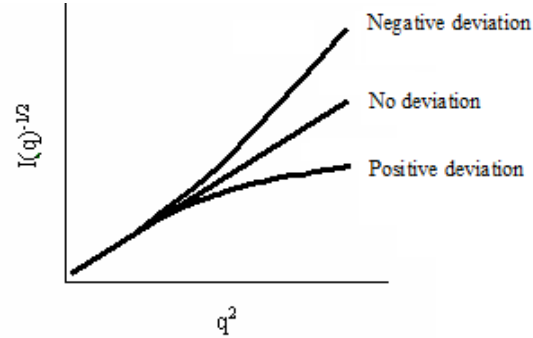

Figure 2  Debye plot with point-collimation in SAXS

$$I(q) = \frac{C}{\left(1 + A_c^2 q^2\right)^2} \quad \text{or} \quad I(q)^{-1/2} = C^{-1/2} + C^{-1/2} A_c^2 q^2 \tag{13}$$

where C is a constant. This equation means that the plot of $I(q)^{-1/2}$ vs $q^2$ appears linear within the entire measured scattering vector range as the middle straight line in Figure 2.

The specific surface can also be calculated with the Debye method [14,15]:

$$S_V = \frac{4P(1-P)}{A_c} \tag{14}$$

If the sample is a porous system, the average size of pores $L_P$ and the average thickness of skeleton $L_S$ can be estimated with the Debye method [15,18]:

$$L_p = \frac{A_c}{1-P} \tag{15}$$

$$L_s = \frac{A_c}{P} \tag{16}$$

**(2) Negative deviation**

For a quasi two-phase system with a diffuse phase boundary or a transition zone (i.e. an interface layer) with average thickness E between two phases, there is a negative deviation from the Debye's theory (see figure 2) [17]. In this case:

$$I(q) = \frac{C}{(1+A_c^2 q^2)^2}\exp(-\sigma^2 q^2) \quad \text{or} \quad \exp\left(-\frac{1}{2}\sigma^2 q^2\right)I(q)^{-1/2} = C^{-1/2} + C^{-1/2}A_c^2 q^2$$

(17)

The physical meaning of $\exp(-\sigma^2 q^2)$ here is the same as that in Equation (5). Using Equation (17) to fit the negative deviation from the Debye's theory, one can derive the constant C and the parameters $\sigma$ and $A_c$. The average thickness E of the interface layer can then be determined from $\sigma$ with Equation (6).

**(3) Positive deviation**

For a quasi two-phase system having sharply defined phase boundaries but with micro-fluctuations of electron density within any phase of two-phase system, the overall scattering displays a positive deviation from the Debye's theory (see Figure 2) [19], and:

$$I(q) = \frac{C}{(1+A_c^2 q^2)^2}\exp(bq^2) \quad \text{or} \quad \exp\left(\frac{1}{2}bq^2\right)I(q)^{-1/2} = C^{-1/2} + C^{-1/2}A_c^2 q^2 \quad (18)$$

The physical meaning of $\exp(bq^2)$ here is the same as in Equation (7).

**(4) Correction of deviation**

Once parameters $\sigma$ and b are derived from the Debye analysis, the correction of negative deviation [20] and positive deviation [19] from the Debye's theory can be performed with Equation (8) and Equation (9), respectively.

**3) Guinier approximation**

Regardless of scatterer shape, the Guinier formula holds at small angle [21]

$$I(q) = I(0) \cdot e^{-\frac{R_G^2 q^2}{3}} \quad \text{or} \quad \ln I(q) = \ln I(0) - \frac{R_G^2}{3} q^2 \tag{19}$$

where the "radius of gyration" $R_G$ is defined as the mean square distance of the scatterers from the centre of their distribution, I(0) is the forward scattering intensity. The Guinier formula is valid assuming that the scatterers are monodisperse, identical, isotropic (but can be extended to any anisotropic form), dilute (so that interference between the various particles is negligible) and randomly oriented; $qR_G \leq 1.3$ (spherical particle); $qR_G \leq 0.7$ (rod-like particle); matrix or solvent scattering is removed. The radius of gyration is the only precise parameter which can be determined by SAXS without invoking supplementary hypotheses.

**4) Shape evaluation**

For a monodisperse system in SAXS, the form factor f of the scatterer is defined as the value of the scatterer volume $V_P$ divided by the spherical scatterer volume $V_S$ having the same radius of gyration $R_G$ as that of the scatterer [8].

$$f = \frac{V_P}{V_S} \tag{20}$$

$$V_P = \frac{2\pi^2 I(0)}{Q} \tag{21}$$

$$V_S = \frac{4\pi}{3} \left( \sqrt{\frac{5}{3}} R_G \right)^3 \tag{22}$$

where the forward scattering intensity I(0) and the radius of gyration $R_G$ can be derived from Equation (19), Q can be computed with Equation (4).

If the scatterer approximates an ellipsoid with axis ratio $\gamma$, then the form factor f can be computed [8]:

$$f = \gamma \left( \frac{3}{2 + \gamma^2} \right)^{3/2} \tag{23}$$

Form factors are easily calculated for some basic shapes such as spheres, cylinders, disks, rods, micelles, lamellas or ellipsoid [22]. The scattering can be interpreted by modeling the scatterer shape, calculating the expected data, and refining the model such that the calculated data best match the experimental curve. Of course shape determination often requires supplementary knowledge from crystallography, nuclear magnetic resonance, or electron microscopy, etc.

In SAXS, the shape of a distance distribution function p(r) can allow qualitative shape evaluation. p(r) is related to the frequency of certain distances r within a scatterer. Therefore it starts from zero at r=0 and goes to zero at the largest diameter r=$r_{max}$ of the scatterer [23]. The shape of the p(r) function already tells something about the shape of the scatterer. If the function is very symmetric, the scatterer is also highly symmetric, like a sphere. p(r) is sensitive to the spatial extent and shape of the monodisperse scatterer and is a useful tool for visibly detecting conformational changes within a macromolecule. p(r) is given by the Fourier transform of the measured scattered intensity I(q) [23]:

$$p(r) = \frac{1}{2\pi^2} \int_0^\infty I(q) qr \sin(qr) dq \qquad (24)$$

The radius of gyration $R_G$ of the scatterer can also be computed from the p(r) [23]:

$$R_G^{\ 2} = \frac{\int_0^{r_{max}} p(r) r^2 dr}{2 \int_0^{r_{max}} p(r) dr} \qquad (25)$$

**5) Scatterers size distribution determination**

SAXS is a valuable non-destructive method for accessing the distribution of nanoparticles or nanopores by size. The scattered intensity I(q) for a polydisperse system of non-interacting particles or pores with the same shape and electron density, but different size can be expressed by [1,24]:

$$I(q) = C \cdot \int_0^\infty D_V(r) \cdot r^3 \cdot I_0(q,r) \cdot dr \qquad (26)$$

where C is a constant. $D_V(r)$ denotes the volume-weighted size distribution or volume distribution of the scatterer with size r. $I_0(q,r)$ is the scattering intensity of the radially symmetric scatterer of size r, with its forward scattering normalized to unity. $I_0(q,r)$ is also often referred to as scatterer form factor P(q).

Different methods have been adopted to compute the size distribution $D_V(r)$ vs r of scatterers from Equation (26) in the literature [1,8]. The optional methods implemented in the program include cascade tangent [1], Maxwellian distribution [1,25], log-normal distribution [1,26], and maximum entropy [27]. The scatterers mean size can then be computed:

$$\bar{r} = \sum r_i D_{V\,i} \qquad (27)$$

**6) Fractal**

The interpretation of the scattering effect from complex, disordered materials notably simplifies when fractal geometry can be applied in the description of their structure. The SAXS intensity from fractal objects has a simple power-law form [28]:

$$I(q) = I_0 q^{-\alpha} \qquad (28)$$

where $I_0$ and $\alpha$ are constants.

The values of the power-law exponent $\alpha$ can be determined from the slope of linear parts of $\log I(q)$ vs. $\log q$ plots. From these values, the mass ($D_m$), pore ($D_p$) and surface ($D_s$) fractal dimensions can be calculated.

For volume (mass or pore) fractals:

$$\alpha = D_m \text{ or } D_p, \text{ so } 1 < \alpha < 3 \qquad (29)$$

whereas for surface fractals:

$$\alpha = 6 - D_s, \text{ so } 3 < \alpha < 4 \qquad (30)$$

The values of the exponent α for both the volume and surface fractal are different. This value may thus be used to distinguish whether the structure of scatterers is a volume or surface fractal.

The self-similarity of real, physical objects can be satisfied only in a statistical sense, and most real objects can reveal fractal properties over a limited range of length:

$$\alpha < l < \xi \tag{31}$$

The ratio $\xi/\alpha$ can be regarded as the measure of fractal range. The values of $\alpha$ and $\xi$ correspond to the scale of the radius of gyration.

## III. Program structure and function

The routines of the program package perform tasks that typically arise during or after data acquisition. The basic structure and functions of the program are schematically displayed in Figure 3. The program is composed of two modules.

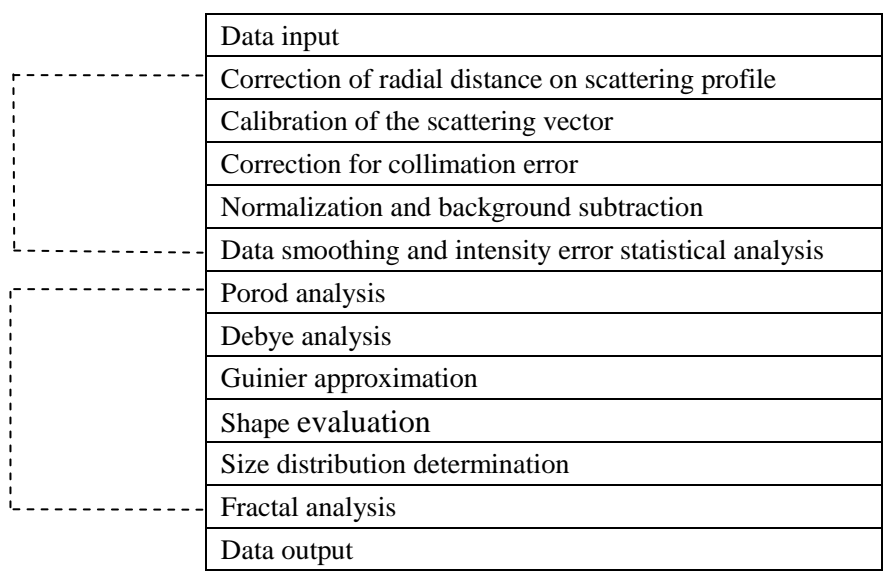

Figure 3  Flow chart of the program S.exe for SAXS data processing and analysis

The first module (I) is used for primary data processing as the correction of the position of the centre of the direct beam, the calculation of the radial distance on the scattering profile, the choice of useful data, the calibration of the scattering vector, the correction for the collimation error, the normalization and background subtraction. The input file of module (I) is the raw 1D data directly recorded by a 1D detector or transformed from the image of a 2D detector. For 2D area detectors, the program does not need any computation of scattering vector or angle during the conversion from 2D to 1D scattering data with other tool (e.g. Fit2D [29]) but directly derives 1D intensity (count) vs relative radial position (distance or pixel or channel) data in either binary or ASCII format. The detector can be set to the direct beam with any angle, and the direct beam can irradiate on or out of the detector according to the experimental needs. After inputting a 1D data file of a standard sample, or a sample or a background, the module (I) firstly checks whether the beamstop shadow lies on the detector and the scattering profile passes approximately through the centre of the beamstop shadow. If so, the module (I) automatically corrects the position of the centre of the direct beam (see Figure 4), calculates the radial distance on the scattering profile and deletes useless data. Otherwise, module (I) uses the diffraction peaks data of a standard sample to create a calibration file named "calibration.dat" to be called later. Then the module (I) performs a series of operation such as conversion of radial distance to scattering vector based on either the standard sample or the knowledge of incident X-ray wavelength λ and sample to detector distance L, correction of the collimation error based on a slit function if necessary [1,5], normalization with either the reading of ion chamber (or photodiode) located downstream the sample or the time of exposure, background subtraction, data smoothing and intensity error statistical analysis [30]. These processing steps are done by calling the corresponding subprograms. The resultant output file with ASCII format of module (I) is about background subtracted scattering data of scattering vector (1/nm) / intensity (a.u.) / error (a.u.) of sample. The data can be used to study sample structure with other modules or other programs.

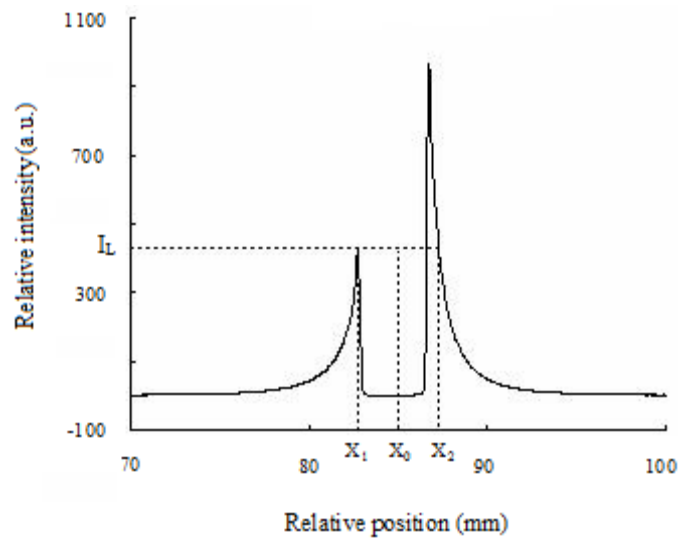

Figure 4  Schematic diagram for correcting the position of direct beam center on an example SAXS profile of a $SiO_2$ colloid sample

The second module (II) is used for data analysis to compute structural parameters of the sample such as scatterer shape, scatterers size distribution, specific surface, porosity, interface layer thickness, correlation distance, fractal dimension, etc. A series of subprograms are integrated in the module (II) to compute different structural parameters for different samples with either the desmeared or slit-smeared data. The main function of this module includes Porod analysis, Debye analysis, Guinier approximation, shape evaluation, size distribution determination and fractal analysis. The results from module (II) containing the structural information of the sample are saved in a file with ASCII format.

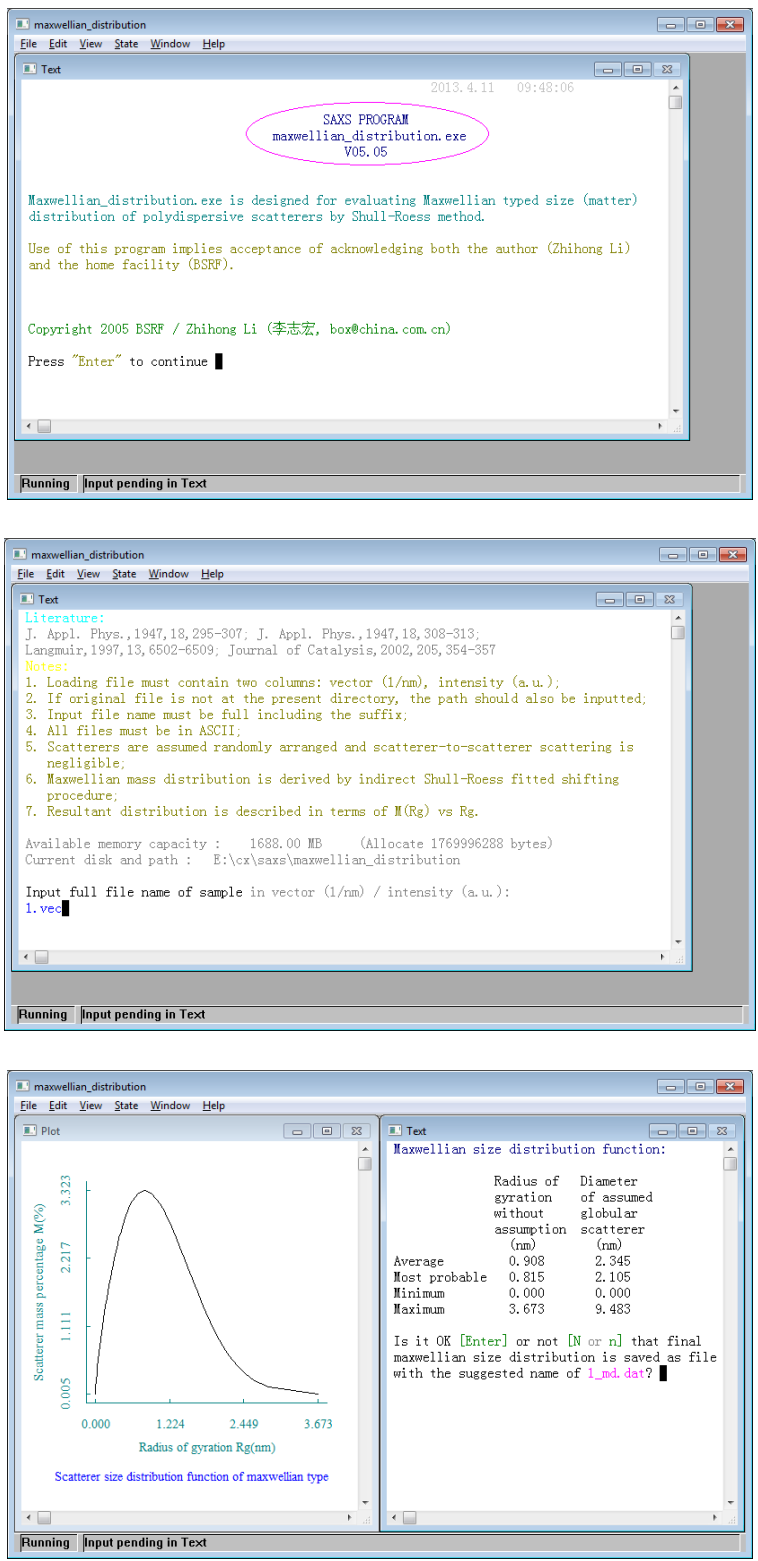

Figure 5: Example of the program execution interface for computation of the scatterers size

distribution by calling the subprogram s3_7 in Module III. The top is about the function introduction, the middle is about the literature, notes and input and the bottom is about the results.

This program is particularly useful with a standard sample for calibration especially when the beam is not perfectly stable. The advantage of this program is to obviate the need of an accurate knowledge of the wavelength and sample-detector distance. The program can be run fully automatically or semi-automatically. It is easy to run the program profiting from the integrated instruction and the step by step operating design (see Figure 5). Each module and each subprogram with special function can be called independently. The result of each module is saved separately, which can be viewed and called freely.

Furthermore, the program is freely available registered, used and updated and is continuously improved.

**Acknowledgement**

The author wishes to thank Dr. Michel Koch (EMBL, Hamburg) for his valuable advice. This study was supported by National Natural Science Foundation of China under Grant No.11079041, No.10835008, No.21176255, No.10979005, No.10979076 and No.111107903928).